%% file: colm2026_conference.tex
\documentclass{article} 
\usepackage[preprint]{colm2026_conference}

\usepackage{microtype}
\usepackage{hyperref}
\usepackage{url}
\usepackage{booktabs}

\usepackage{graphicx}
\usepackage{xspace}
\usepackage{caption}
\usepackage{subcaption}
\usepackage{multirow}
\usepackage{algorithm}
\usepackage[noend]{algpseudocode}

\usepackage[utf8]{inputenc} 
\usepackage[T1]{fontenc}    
\usepackage{amsfonts}       
\usepackage{nicefrac}       
\usepackage{xcolor}         
\usepackage{amsmath}
\usepackage{colortbl}
\usepackage{soul}
\usepackage{bm}
\usepackage{tcolorbox}
\usepackage{cleveref}
\newtcolorbox{prompt}[1]{colback=gray!20,colframe=gray!50!black,fonttitle=\bfseries,title=#1}

\newcommand{\ours}{{RAR}\xspace}
\newcommand{\markblue}[1]{\textcolor{black}{#1}}

\usepackage{lineno}

\definecolor{darkblue}{rgb}{0, 0, 0.5}
\hypersetup{colorlinks=true, citecolor=darkblue, linkcolor=darkblue, urlcolor=darkblue}

\title{Retrieval Augmented Conversational Recommendation with Reinforcement Learning}

\author{Zhenrui Yue$^1$, Honglei Zhuang$^2$, Zhen Qin$^2$, Zhankui He$^3$, Huimin Zeng$^1$, \\
\textbf{Julian McAuley$^3$, Dong Wang$^1$} \\
$^1$University of Illinois Urbana-Champaign, $^2$Google DeepMind, $^3$UC San Diego \\
\texttt{\{zhenrui3, dwang24\}@illinois.edu} \\
}

\begin{document}

\ifcolmsubmission
\linenumbers
\fi

\maketitle

\begin{abstract}
\markblue{
Large language models (LLMs) exhibit enhanced capabilities in language understanding and generation. By utilizing their embedded knowledge, LLMs are increasingly used as conversational recommender systems (CRS), achieving improved recommendation performance across diverse scenarios. However, existing LLM-based methods rely on pretrained knowledge without external retrieval mechanisms for novel items. Additionally, the lack of a unified corpus poses challenges for integrating retrieval augmentation into CRS. Motivated by these challenges, we present \ours, a novel two-stage \ul{r}etrieval \ul{a}ugmented conversational \ul{r}ecommendation framework that aligns retrieval and generation to enhance both performance and factuality. To support this framework and provide a unified corpus, we construct a large-scale movie corpus, comprising over 300k movies with rich metadata, such as titles, casts and plot summaries. Leveraging this data, our primary contribution is \ours, the first framework to departs from standard two-stage CRS by dynamically bridging retrieval and generation. First, a retriever model generates candidate items based on user history; in the subsequent stage, an LLM refines the recommendations by incorporating conversational context with retrieved results. In addition, we introduce a novel reinforcement learning (RL) method that leverages LLM feedback to iteratively update the retriever. By creating a collaborative feedback loop that reinforces sampled candidate sets with higher ranking metrics, \ours effectively mitigates the misalignment between the retrieval and generation stages. Furthermore, grounding the LLM in factual metadata allows our RL-driven approach to capture subtle user intentions and generate context-aware recommendations with reduced hallucinations. We validate our approach through extensive experiments on multiple CRS benchmarks, where \ours consistently outperforms state-of-the-art baseline methods.
}
\end{abstract}

\input{latex/1_intro}
\input{latex/2_related}
\input{latex/3_method}
\input{latex/4_exp}
\input{latex/5_conclusion}

\bibliography{custom, anthology}
\bibliographystyle{colm2026_conference}

\appendix

\input{latex/6_appendix}

\end{document}

%% file: latex/1_intro.tex
\section{Introduction}

Conversational recommender systems (CRS) have emerged as a promising paradigm for providing personalized recommendations through natural language interactions~\citep{zhang2018towards, li2018towards, kang-etal-2019-recommendation, hayati2020inspired, gao2023chat, he2023large}. Recent advancements in large language models (LLM)~\citep{achiam2023gpt, dubey2024llama, comanici2025gemini} have further enabled CRS methods to utilize their extensive knowledge and diverse generation capabilities, demonstrating improvements across different recommendation scenarios~\citep{feng2023large, yang2024unleashing, li2024incorporating, hui2026toward}. For example, \citet{he2023large} adopt LLMs to generate movie candidates and achieve superior performance compared to traditional CRS methods.

\begin{figure}[t]
    \centering
    \includegraphics[trim=3.7cm 5.2cm 3.9cm 5.8cm, clip, width=0.95\textwidth]{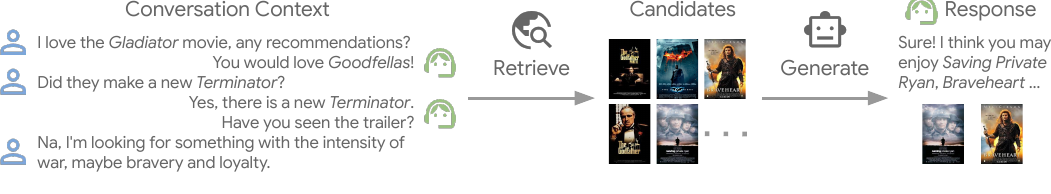}
    \caption{Our retrieval-augmented conversational recommendation framework, where a retriever gathers candidate items and the LLM generates the response conditioned on them.}
    \label{fig:intro}
\end{figure}

Nevertheless, language models often lack the essential knowledge to recommend the most suitable items (e.g., being unaware of relevant options)~\citep{he2023large, yang2024unleashing, wu2024coral}. As a solution, traditional CRS methods exploit search modules or knowledge graphs to provide additional contextual information~\citep{zhang2018towards, chen-etal-2019-towards, zhou2020improving}. Similarly, LLM-based CRS methods leverage knowledge graph retrieval or training on a broad range of items to improve recommendation relevance~\citep{li2024incorporating, jeon2025adapting, zare2025conversational}. A notable example is ReFICR, where LLMs are trained on extensive curated data to perform sub-tasks including indexing, retrieval and generation\citep{yang2024unleashing}. In \Cref{fig:intro}, we illustrate this general approach of the two-stage retrieval augmented conversational recommendation.

While LLMs can accommodate novel items through additional training, scaling this process to extensive items and conversations is often infeasible, particularly when managing large volumes of items and noisy conversations~\citep{jannach2021survey, zhu2025llm, surana2025reviews}. On the other hand, retrieving from knowledge graph requires significant efforts in data preprocessing, modeling and graph indexing. Graph retrieval can also be more computationally intensive when indexing or traversing through large-scale graphs~\citep{li2024incorporating, peng2024graph}. Although embedding-based retrieval is a straightforward alternative~\citep{reimers-gurevych-2019-sentence, wang2022text, chen-etal-2024-m3}, its effectiveness is limited by the lack of a unified corpus that provides comprehensive attribute-level metadata (e.g., plot, cast) corpus. As such, embedding-based retrieval remains largely under-explored in conversational recommendation. Furthermore, LLM-based CRS often exhibits retrieval–generation misalignment: when the retriever returns sub-optimal candidates (e.g., low-relevance items), LLM could amplify such deficiencies and cause deteriorated accuracy on cold-start or less-popular items~\citep{he2023large, kemper2024retrieval}.

In this paper, we investigate how embedding-based retrieval augmentation can be used to improve two-stage conversational recommendation. To address the limitations of existing, small-scale movie corpora (e.g., REDIAL with $\sim$7k titles), we curate a comprehensive evaluation benchmark of over 300k movies, systematically enriched with metadata (e.g., titles, casts, and plot summaries). By establishing this robust, large-scale foundation, our primary contribution is \ours, a novel LLM-based \ul{r}etrieval \ul{a}ugmented conversational \ul{r}ecommendation framework. In the first stage, we employ a retriever model to select candidate items based on the conversational history. Subsequently, an LLM refines the recommendations by incorporating conversational context with retrieved candidates, enabling personalized and fine-grained recommendation results. To enhance retrieval-generation alignment, we further optimize the retriever by leveraging LLM feedback for reinforcement learning (RL). This feedback enables \ours to iteratively update the retriever through online, on-policy preference optimization. As a result, \ours can be adapted for any black-box LLMs and deliver high-quality, context-aware recommendations. We demonstrate the efficacy of \ours with extensive experiments, where RAR consistently outperforms baselines on multiple benchmark datasets. We summarize our contributions in the following\footnote{Our data and code can be accessed at https://github.com/Yueeeeeeee/RAR.}:
\begin{enumerate}
    \item We present the first large-scale study of embedding-based retrieval augmentation for conversational recommendation. To support this scale, we synthesize a highly-enriched, structured baseline of over 300k films with comprehensive metadata.
    \item We propose \ours, an LLM-based CRS framework with two-stage retrieval augmentation. By utilizing LLM feedback as reward signals, we design an online, on-policy reinforcement learning framework that enhances retrieval–generation alignment.
    \item We demonstrate the effectiveness our approach by utilizing our curated retrieval corpus, where the RL post-trained \ours consistently outperforms state-of-the-art baseline methods, achieving considerable improvements in recommendation performance on established benchmark datasets.
\end{enumerate}

%% file: latex/2_related.tex
\section{Related Work}

\subsection{Retrieval Augmented Generation}
Retrieval augmented generation (RAG) enhances language modeling by integrating external knowledge into context~\citep{lewis2020retrieval, guu2020retrieval, karpukhin2020dense}. Beyond na\"ive RAG, performance is improved by refining the retriever (e.g., via joint optimization~\citep{lin2024ra})~\citep{trivedi2023interleaving, jiang2023active, shi2024replug, sarthi2024raptor}, upgrading document encoding~\citep{izacard2021leveraging, borgeaud2022improving, izacard2023atlas}, and selectively filtering retrieved knowledge to minimize irrelevant context~\citep{yoran2023making, yan2024corrective, yue2024inference, jin2025search}. While concurrently applied to sequential recommendation~\citep{wu2024coral, kemper2024retrieval}, RAG's potential for conversational recommendation remains largely under-explored. To bridge this gap, we build a unified movie metadata corpus and examine retrieval-augmented LLMs, aiming to enhance CRS performance via effective knowledge retrieval.

\subsection{Conversational Recommendation}
Conversational recommender systems (CRS) capture user preferences via multi-turn interactions to improve recommendations. Early methods relied on complex, multi-module systems for dialogue and recommendation~\citep{zhang2018towards, li2018towards, kang-etal-2019-recommendation, hayati2020inspired, lei2020estimation}. Later, language modeling and knowledge integration enabled better contextualization of user intent~\citep{chen-etal-2019-towards, zhou2020improving, lu-etal-2021-revcore, wang2022towards}, such as inferring preferences from incomplete knowledge graphs~\citep{zhang2023variational}. Recently, large language models (LLMs)~\citep{achiam2023gpt, reid2024gemini, dubey2024llama} have transformed CRS through advanced dialogue capabilities and precise preference modeling~\citep{gao2023chat, he2023large, li2024incorporating, he2024reindex, yang2024unleashing, hui2026toward}, including using LLMs for user simulation~\citep{zhu2025llm}. Despite these advances, integrating LLMs into CRS and improving the retrieval–generation alignment remain under-investigated.

\subsection{Reinforcement Learning}
Reinforcement learning (RL) maximizes cumulative rewards through environmental interaction~\citep{sutton1998reinforcement}. RL has been adapted to align LLMs via human feedback (RLHF)~\citep{ouyang2022training}, typically utilizing policy gradient variants~\citep{sutton1999policy}. Methods like A2C and PPO~\citep{mnih2016asynchronous, schulman2017proximal} leverage learned baselines and clipped surrogate objectives for stable training. Alternatively, Direct Preference Optimization (DPO)~\citep{rafailov2023direct} efficiently optimizes offline pairwise preferences, though online RL methods consistently achieve superior performance~\citep{xu2024dpo}. Recent online advances like GRPO evaluate candidates against a group-based baseline, significantly reducing memory overhead without sacrificing stability~\citep{shao2024deepseekmath, hu2025reinforce++, yue2025hybrid, yue2026dr, hubotter2026reinforcement}. In this work, we propose a novel online RL-driven approach for two-stage conversational recommender systems (CRS) and empirically compare different RL algorithms to identify the optimal strategy for recommendation.

%% file: latex/3_method.tex
\section{Methodology}

\subsection{Setup}
Our conversational recommendation framework is based on a two-stage setting with both retrieval and generation. Consider a conversation $\mathcal{C} = (r_t, s_t, I_t)_{t=1}^{T}$ consisting of $T$ turns, $r_t$, $s_t$ and $I_t$ refer to the role (i.e., \emph{seeker} or \emph{recommender}), sequence (conversation turn) and mentioned items. Each element in $I_t$ should be included in the corpus $\mathcal{I}$, namely $\forall i \in I_t, \; i \in \mathcal{I}$. Typically, the \emph{seeker} and \emph{recommender} take turns to converse in $\mathcal{C}$ until the seeker finds desired item(s). Following previous research~\citep{li2018towards, chen-etal-2019-towards, wang2022towards, he2023large}, the conversational recommender aims to generate a ranked list $\hat{I}_t$ for the $t$-th turn when $r_t$ is the \emph{recommender}, such that the generated $\hat{I}_t$ best matches the ground truth items $I_t$. Specifically for our two-stage framework:
\begin{itemize}
    \item \emph{Retrieval}: In the first stage, the retriever model $f_{\mathrm{ret}}$ uses the previous items as query to select an initial candidate set $C_t$, i.e., $C_t = f_{\mathrm{ret}}(\{ I_{\tau} \}_{\tau=1}^{t-1})$.
    \item \emph{Generation}: Then, we leverage a LLM $f_{\mathrm{llm}}$ to generate $\hat{I}_t$ upon user conversation history and retrieved items. Formally, this is expressed as $\hat{I}_t = f_{\mathrm{llm}}(\{ s_k \}_{k=1}^{t-1}, C_t)$, enabling an in-depth analysis of user preferences and potential interests.
\end{itemize}

To fully leverage black-box LLMs without excessive training costs, we refrain from directly training $f_{\mathrm{llm}}$ and instead focus on optimizing the retriever model $f_{\mathrm{ret}}$. This design choice offers two key advantages: (1)~\ours can be combined with any LLM choice, regardless of whether the LLM is open- or closed-source; (2)~by enhancing $f_{\mathrm{ret}}$, we augment $f_{\mathrm{llm}}$ through the incorporation of up-to-date external knowledge (e.g., novel items). Our framework optimizes $f_{\mathrm{ret}}$ (parameterized by $\theta$) to maximize the expected reward. Since the retriever and generator are not end-to-end differentiable, generator errors cannot be directly backpropagated. We therefore train $f_{\mathrm{ret}}$ via reinforcement learning, maximizing reward $r$ over samples from the retriever across conversations ($\mathcal{C}$) and timesteps ($t$) in dataset $\mathcal{X}$:
\begin{equation}
   \max_{\theta} \mathbb{E}_{\mathcal{C} \sim \mathcal{X}, t \sim \{1, \ldots, T(\mathcal{C}) \}, C_t \sim f_{\mathrm{ret}}(\{ I_{\tau} \}_{\tau=1}^{t-1})} [r(f_{\mathrm{llm}}(\{ s_k \}_{k=1}^{t-1}, C_t), I_t)],
\end{equation}
where $r$ is a ranking-based reward for the sampled candidate set $C_t$ calculated based on the output of $f_{\mathrm{llm}}$ (e.g., NDCG scores). That is, the post-training process updates $\theta$ by maximizing the ranking reward, thereby aligning the retrieval and ranking stages in \ours.

\subsection{Corpus Construction}
\label{sec:corpus}
Existing conversational recommendation datasets either rely on large-scale knowledge graphs or on small, domain-specific corpus~\citep{li2018towards, hayati2020inspired, he2023large}. Therefore, embedding-based retrieval is often impractical with existing data. To address this limitation, we collect and curate a unified text corpus for movie recommendation across datasets. Our initial corpus is built from all available entries across multiple sources, including IMDb genre, IMDb media and Inspired~\citep{hayati2020inspired, rajugc_imdbmovies, brightdata_imdbmedia}. Next, we augment the corpus with inferred entries from MovieLens, Redial and Reddit~\citep{harper2015movielens, li2018towards, he2023large}. Then, we aggregate duplicate film entries by cross-referencing and selecting the most reliable metadata. Additionally, we collect missing metadata from online resources for incomplete entries, we also map items in the corpus to the corresponding entries in the datasets (i.e., MovieLens, Inspired, Redial and Reddit). That is, we perform entity recognition to identify each movie mentioned in the conversation and map it to its corresponding entry in the corpus. Finally, we remove entries that were either incomplete or could not be collected, we report detailed information on corpus collection and preprocessing in \Cref{sec:app}. In summary, we obtain 337,731 entries with comprehensive metadata, focusing primarily on English-language films (see example in \Cref{tab:metadata-example}). The corpus spans from as early as 1888 to upcoming releases in 2029, accounting for approximately half of all movies listed on IMDb as of 2025. Leveraging the collected corpus and constructed correspondence, we pretrain a retriever model on MovieLens by incorporating recommendation data along with negative examples sampled from the corpus~\citep{harper2015movielens, yue2024linear}. Next, the retriever is optimized with sequential patterns derived from the conversation data, we introduce the details in the following.

\begin{figure*}[t]
    \centering
    \includegraphics[trim=0.9cm 4.2cm 1.1cm 4.5cm, clip, width=0.95\textwidth]{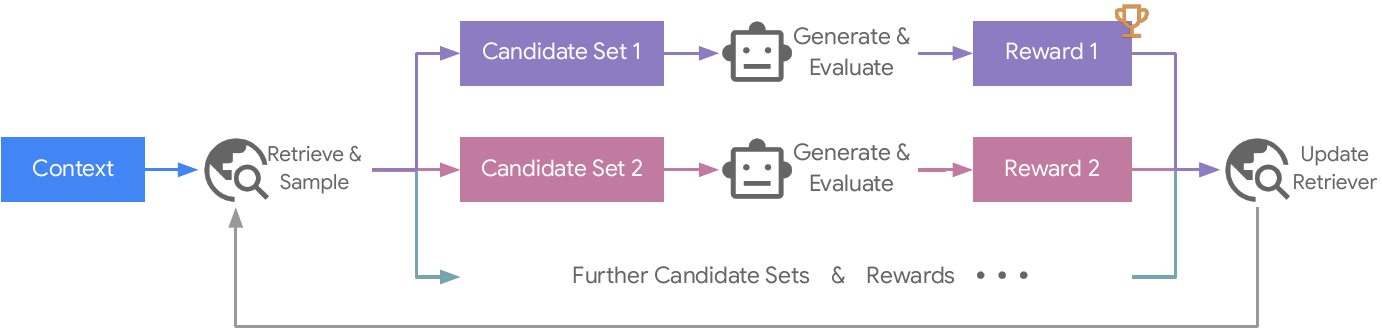}
    \caption{The proposed online, on-policy preference optimization in \ours iteratively refines the retriever model by sampling candidate sets, collecting LLM feedback, and then updating the retriever through reinforcement learning.}
    \label{fig:method}
\end{figure*}

\subsection{The proposed \ours}
\textbf{Retriever.}
For \ours, we adopt the linear recurrent units for sequential recommendation (LRURec) as $f_{\mathrm{ret}}$, with a comparison of different models in \Cref{sec:exp}. LRURec utilizes state space modeling (SSM) to efficiently train and infer on sequential input: 
\begin{equation}
    h_t = A h_{t-1} + B e_t, \quad o_t = C h_t + I e_t,
    \label{eq:linear-recurrence}
\end{equation}
with $A$, $B$, $C$ being matrices of shape $\mathbb{R}^{H \times H}$, and $I$ is the identity matrix. $h$, $e$, $o$ denote the hidden state, item embedding and output hidden state. By leveraging LRURec's linearity via parallel scan, we reduce time complexity to $\mathcal{O}(\mathrm{log}(t))$, significantly enhancing training and inference efficiency. To encode the corpus, we leverage Qwen~3 to build the embedding table~\citep{zhang2025qwen3}. The output state $o_t$ is used to compute similarity scores for all corpus entries, and the top-$k$ items are retrieved and stored as $C_t$ for the generation stage.

\noindent
\textbf{Generator.}
In the generation stage, our primary objective is to produce refined recommendations $\hat{I}_t$ by integrating the conversation context $\{ s_k \}_{k=1}^{t-1}$ and the retrieved items $C_t$. To achieve this, we employ a black-box LLM $f_{\mathrm{llm}}$ and construct an input prompt that combines clear instructions, detailed item metadata, and the conversation history. The LLM is instructed to generate $k$ recommendations based on this comprehensive context. Further details of our prompt design are provided in \Cref{sec:app}. For each conversation, we divide it into sub-conversations by splitting at turns where the role $r_t$ is designated as \emph{recommender}. We also exclude ground truth items that have already appeared in earlier conversation turns to avoid shortcut learning~\citep{he2023large, geirhos2020shortcut}. During inference, we retrieve relevant items from the corpus to construct the prompt. The generator $f_{\mathrm{llm}}$ then produces recommendations, which are post-processed to yield the final ranked list $\hat{I}_t$. Unlike prior works~\citep{he2023large, kemper2024retrieval}, we pair a simple retriever with a black-box LLM to inject external, up-to-date knowledge into the generation stage, enabling the LLM to access novel items for fine-grained, context-aware recommendations.

\noindent
\textbf{Retriever Preference Optimization.}
While \ours provides a retrieval augmented generation framework for conversational recommendation, these two stages are not jointly trained for optimal alignment. Yet overlooking the dynamics between both stages can result in a sub-optimal solution~\citep{ma2020off, higley2022building, lin2024ra}. We thus propose an online, on-policy preference optimization method that leverages real-time feedback from the LLM generator $f_{\mathrm{llm}}$. This allows \ours to iteratively sample candidate sets from the retriever and utilize LLM feedback to optimize current policy (see \Cref{fig:method}). In our setup, we consider a frozen generator $f_{\mathrm{llm}}$ and focus on post-training the retriever (parametrized by $\theta$). This approach enables efficient optimization of the smaller $f_{\mathrm{ret}}$ to achieve improved retrieval. Specifically for candidate set $C_t = \{ c_i \}_{i=1}^{k}$ sampled from $\pi_{\theta}$, we define the likelihood of $C_t$ given history items $\{ I_{\tau} \}_{\tau=1}^{t-1}$ using the Plackett-Luce model~\citep{plackett1975analysis}:
\begin{equation}
    P_{\theta}(C_t | \{ I_{\tau} \}_{\tau=1}^{t-1}) = \prod_{i=1}^{k} \frac{\exp(s_{\sigma(i)})}{\sum_{j \in \mathcal{I} \setminus \{\sigma(1), \dots, \sigma(i-1)\}} \exp(s_j)},
    \label{eq:likelihood}
\end{equation}
where $\sigma$ is the permutation that ranks items by descending retriever score, and $s_{\sigma(i)}$ denotes the score assigned by $f_{\mathrm{ret}}$ to the $i$-th ranked item. This formulation treats the candidate set as a sequential selection process without replacement: at each step, an item is drawn from the remaining pool with probability proportional to its retriever score. For each turn, we sample multiple candidate sets and use $f_{\mathrm{llm}}$ to rank and compute rewards. We then apply policy gradients on the resulting advantages and log-likelihood to optimize the retriever, effectively enhancing the synergy between the retrieval and ranking stages.

\textbf{Pairwise and Multi-Sample Reinforcement Learning}.
For pairwise RL, we adopt an online, on-policy DPO approach to maximize the probability of retrieving a favored candidate set $C_w$ over a disfavored set $C_l$. Because $f_{\mathrm{llm}}$ acts as a black-box generator, we compute the final reward as the NDCG ranking score by identifying the rank of the ground truth items directly from the LLM's output. At each timestep, we sample two candidate sets of size $k$ from policy $\pi_{\theta}$; the set that yields the higher NDCG score is designated $C_w$. In other words, the candidate set where the label item achieves a higher rank is considered preferable. Given history $\{ I_{\tau} \}_{\tau=1}^{t-1}$ and candidate sets, we minimize the preference loss $\mathcal{L}_{\mathrm{dpo}}$:
\begin{equation}
    \mathcal{L}_{\mathrm{dpo}} = -\log \sigma ( \beta \log \frac{\pi_{\theta} (C_w | \{ I_{\tau} \}_{\tau=1}^{t-1})}{\pi_{\mathrm{ref}} (C_w | \{ I_{\tau} \}_{\tau=1}^{t-1})} - \beta \log \frac{\pi_{\theta} (C_l | \{ I_{\tau} \}_{\tau=1}^{t-1})}{\pi_{\mathrm{ref}} (C_l | \{ I_{\tau} \}_{\tau=1}^{t-1})} )
\label{eq:dpo_loss}
\end{equation}
where $\beta$ controls the strength of preference learning, while the reference model $\pi_{\mathrm{ref}}$ constrains the policy updates. This objective is designed to increase the relative probability of sampling $C_w$ over $C_l$ while preventing excessive policy divergence. In addition to online DPO, we further introduce a multi-sample RL approach based on GRPO~\citep{shao2024deepseekmath}. GRPO samples a group of $g$ candidate sets and computes the advantages by standardizing the rewards within the group (i.e., $\hat{A}_i = \frac{r_i - \texttt{mean}([r_1, r_2, \ldots, r_g])}{\texttt{std}([r_1, r_2, \ldots, r_g])}$):
\begin{equation}
\mathcal{L}_{\mathrm{grpo}} = \mathbb{E}_{\mathcal{C} \sim \mathcal{X}, t \sim \{1, \ldots, T(\mathcal{C}) \}, \{ C_i \}_{i=1}^{g} \sim f_{\mathrm{ret}}(\{ I_{\tau} \}_{\tau=1}^{t-1})} \left[ - \frac{1}{g} \sum_{i=1}^{g} \log \pi_{\theta} (C_i | \{ I_{\tau} \}_{\tau=1}^{t-1}) \hat{A}_i \right] +\beta \mathbb{D}_{K L},
\label{eq:grpo_loss}
\end{equation}
where $\beta$ controls the regularization strength. The resulting algorithm, detailed in \Cref{alg:rl}, is lightweight, on-policy and can be seamlessly combined with further optimizations.

\textbf{Training Objective}.
To stabilize reinforcement preference learning and maintain the retriever's standalone quality, we incorporate a supervised negative log-likelihood term, $\mathcal{L}_{\mathrm{nll}}$. Consequently, the full preference optimization objective is:
\begin{equation}
    \mathcal{L} = \mathcal{L}_{\mathrm{nll}} + \mathcal{L}_{\mathrm{rl}},
\end{equation}
where $\mathcal{L}_{\mathrm{rl}}$ represents either the DPO or GRPO loss. While DPO trains the policy to favor preferred candidate sets by maximizing their relative selection probability, GRPO refines policy gradients by calculating advantages across groups of samples to amplify the likelihood of high-rewarding sets. By integrating RL loss with the negative log-likelihood term, \ours bridges the retrieval and generation stages, continually refining $f_{\mathrm{ret}}$ to ensure the delivery of diverse, high-quality candidate sets for conversational recommendation.

%% file: latex/4_exp.tex
\section{Experiments}
\label{sec:exp}

\begin{table*}[t]
\noindent\makebox[\textwidth]{
\resizebox{1.0\textwidth}{!}{
\begin{tabular}{@{}lcccccccccccc@{}}
\toprule
\multirow{2}{*}{Method} & \multicolumn{4}{c}{\textbf{Inspired}}                             & \multicolumn{4}{c}{\textbf{Redial}}                               & \multicolumn{4}{c}{\textbf{Reddit}}                                                                   \\ \cmidrule(l){2-5} \cmidrule(l){6-9} \cmidrule(l){10-13}
                        & N@5            & R@5            & N@10           & R@10           & N@5            & R@5            & N@10           & R@10           & N@5                     & R@5                     & N@10                    & R@10                    \\ \midrule
KBRD                    & .0466          & .0815          & .0472          & .0732          & .0388          & .0582          & .0453          & .078           & .0066                   & .0079                   & .0078                   & .0098                   \\
KGSF                    & .0656          & .0815          & .0673          & .0869          & .0434          & .0672          & .0497          & .0864          & .0155                   & .0172                   & .0155                   & .0172                   \\
UniCRS                  & .0676          & .0927          & .0750          & .1032          & .0425          & .0646          & .0504          & .0887          & .0258                   & .0376                   & .0363                   & .0479                   \\ \midrule
SASRec                  & .0564          & .0870          & .0655          & .1304          & .0558          & .0795          & .0681          & .1176          & .0288                   & .0401                   & .0339                   & .0545                   \\
FMLPRec                 & .0620          & .0815          & .0726          & .1141          & .0504          & .0784          & .0639          & .1123          & .0315                   & .0434                   & .0342                   & .0534                   \\
LRURec                  & .0671          & .0978          & .0793          & .1359          & .0539          & .0771          & .0650          & .1111          & .0316                   & .0430                   & .0346                   & .0522                   \\ \midrule
SFT$_{\text{Qwen}}$     & .0609          & .0938          & .0626          & .0990          & .0454          & .0684          & .0526          & .0907          & .0344                   & .0484                   & .0394                   & .0633                   \\
SFT$_{\text{Gem}}$      & .0859          & .1076          & .1034          & .1544          & .0574          & .0828          & .0662          & .1097          & .0455                   & .0604                   & .0497                   & .0708                   \\
SFT$_{\text{GPT}}$      &  \ul{.0997}    &  \ul{.1214}    &  \ul{.1091}    & .1491          & .0599          & .0887          & .0700          &  \ul{.1197}    & .0489                   & .0651                   &  \ul{.0558}             &  \ul{.0843}             \\ \midrule
RAR$_{\text{Qwen}}$     & .0693          & .0980          & .0773          & .1241          & .0491          & .0704          & .0569          & .0947          & .0368                   & .0536                   & .0444                   & .0770                   \\
RAR$_{\text{Gem}}$      & .0916          & .1145          & .1046          &  \ul{.1587}    &  \ul{.0632}    &  \ul{.0894}    & \textbf{.0721} & .1169          &  \ul{.0502}             &  \ul{.0661}             & .0531                   & .0799                   \\
RAR$_{\text{GPT}}$      & \textbf{.1091} & \textbf{.1422} & \textbf{.1180} & \textbf{.1700} & \textbf{.0620} & \textbf{.0932} &  \ul{.0718}    & \textbf{.1236} & \textit{\textbf{.0551}} & \textit{\textbf{.0716}} & \textit{\textbf{.0593}} & \textit{\textbf{.0846}} \\ \bottomrule
\end{tabular}
}
}
\caption{Main results of the online DPO-based \ours and baseline methods on conversational recommendation. For clarity, the best results for each dataset and metric are highlighted in \textbf{bold}, while the second-best results are \ul{underlined}.}
\label{tab:main}
\end{table*}

\subsection{Experiment Settings}

Our model is evaluated on three widely-used datasets for conversational recommendation: Inspired, Redial and Reddit~\citep{hayati2020inspired, li2018towards, he2023large}. We adopt multiple baselines for comparison, including both \emph{traditional} and \emph{LLM-based} methods. In particular, we adopt \textit{traditional CRS}: KBRD, KGSF and UniCRS~\citep{chen-etal-2019-towards, zhou2020improving, wang2022towards}; \textit{sequential methods}: SASRec, FMLPRec and LRURec~\citep{kang2018self, zhou2022filter, yue2024linear}; \textit{retrieval augmented CRS with supervised fine-tuning} (SFT) methods include Qwen3-8B (Qwen), GPT-5 mini (GPT) and Gemini 3 Flash (Gem)~\citep{yang2025qwen3, singh2025openai}, where we perform SFT on the retriever model. Similarly, we experiment \ours with Qwen, GPT and Gemini as LLM generator. For our retriever model, we compare GRU4Rec, SASRec, FMLPRec and LRURec~\citep{hidasi2015session, kang2018self, zhou2022filter, yue2024linear}. For all methods and datasets, the maximum history length was set to 64 (i.e. history items), and the retrieval size was set to $k=25$ in our main experiments. The evaluation metrics are NDCG and Recall at 5 and 10. For all evaluated methods, we saved the model with the best validation NDCG@10 score. Further training and evaluation details are reported in \Cref{sec:app2}.

\subsection{Experiment Results and Analysis}

\textbf{RQ1. How does \ours perform in conversational recommendation?}
We first discuss the evaluation results for conversational recommendation datasets, as reported in \Cref{tab:main}. Based on the presented results, we have several key observations:
(1) Overall Performance: Across all datasets and metrics, \ours demonstrates a clear and consistent superiority over traditional, sequential, and SFT baselines, highlighting its robust capability to retrieve and generate high-quality recommendations. In particular, \ours achieves an average improvement of 7.60\% over the best baseline results.
(2) Impact of Base LLMs: Among the evaluated models enhanced by \ours, GPT delivers the highest overall performance across the datasets, closely followed by Gemini, and then Qwen. This indicates the importance of the LLMs' inherent capabilities, where GPT and Gemini stand as the state-of-the-art closed-source models, while Qwen presents a powerful open-source alternative for conversational recommendation.
(3) Comparison to Traditional and SFT Baselines: Overall, baseline comparisons reveal that while SFT-based retrieval augmentation outperforms traditional methods, \ours further surpasses standard fine-tuning. It achieves average gains of 11.9\%, 7.5\% and 7.7\% for Qwen, Gemini and GPT backbones over their SFT-only counterparts, respectively, demonstrating the efficacy of our RL-based post-training.
(4) Trends Across Metrics: Although methods exhibit some variability across metrics, the trends are consistent: performance gains of \ours against the best baselines are particularly pronounced on top-$k$ metrics such as N@5 and R@5. For example, the ranking performance improvement for @5 metrics averages 9.33\% compared to 5.86\% for @10 metrics, suggesting that our preference optimization strategy is especially effective at pushing the most relevant items to the top of the list.
Overall, our findings show that \ours effectively integrates retrieval augmentation and RL-based preference optimization to enhance recommendation performance.

\begin{table}[t]
    \centering
    \begin{minipage}{0.49\textwidth}
        \centering
        \resizebox{\linewidth}{!}{
        \small
        \begin{tabular}{@{}lccccc@{}}
        \toprule
        Retriever & & N@10            & R@10            & N@20            & R@20            \\ \midrule
        GRU4Rec   & & .1250          & .2015          & .1351          & .2558          \\
        BERT4Rec  & & .1469          & .2337          & .1598          & .2889          \\
        SASRec    & & .1430          & .2363          & .1637          & .3086          \\
        FMLPRec   & & .1460          & .2411          & .1689          & .3125          \\
        LRURec    & & \textbf{.1483} & \textbf{.2508} & \textbf{.1700} & \textbf{.3365} \\ \bottomrule
        \end{tabular}
        }
        \caption{Different retriever model performance pretrained on MovieLens.}
        \label{tab:retriever}
    \end{minipage}\hfill
    \begin{minipage}{0.49\textwidth}
        \centering
        \resizebox{\linewidth}{!}{
        \small
        \begin{tabular}{@{}llcccc@{}}
        \toprule
        Dataset                   & Method & N@5            & R@5            & N@10           & R@10           \\ \midrule
        \multirow{2}{*}{Inspired} & DPO    & .0693          & .0980          & .0773          & \textbf{.1241} \\ 
                                  & GRPO   & \textbf{.0753} & \textbf{.0997} & \textbf{.0807} & .1162          \\ \midrule
        \multirow{2}{*}{Redial}   & DPO    & .0491          & .0704          & .0569          & .0947          \\ 
                                  & GRPO   & \textbf{.0493} & \textbf{.0706} & \textbf{.0575} & \textbf{.0954} \\ \midrule
        \multirow{2}{*}{Reddit}   & DPO    & .0368          & .0536          & .0444          & \textbf{.0770} \\
                                  & GRPO   & \textbf{.0385} & \textbf{.0547} & \textbf{.0450} & .0765          \\ \bottomrule
        \end{tabular}
        }
        \caption{DPO \& GRPO Comparison on Qwen.}
        \label{tab:dpo-grpo}
    \end{minipage}
    
\end{table}

\begin{figure}[t]
    \centering
    \begin{minipage}{0.48\textwidth}
        \centering
        \resizebox{\linewidth}{!}{
        \small
        \begin{tabular}{@{}lcccc@{}}
        \toprule
        \multirow{2}{*}{Method} & \multicolumn{4}{c}{\textbf{Average Performance}}                      \\ \cmidrule(l){2-5} 
                                & N@5            & R@5            & N@10           & R@10           \\ \midrule
        SimPO$_{\text{Qwen}}$   & .0499          & .0699          & .0575          & .0946          \\
        SimPO$_{\text{Gem}}$    & .0635          & .0825          & .0718          & .1136          \\
        SimPO$_{\text{GPT}}$    & .0727          & .1011          & .0792          & .1203          \\ \midrule
        DPO$_{\text{Qwen}}$     & \textbf{.0517} & \textbf{.0740} & \textbf{.0595} & \textbf{.0986} \\
        DPO$_{\text{Gem}}$      & \textbf{.0683} & \textbf{.0900} & \textbf{.0766} & \textbf{.1185} \\
        DPO$_{\text{GPT}}$      & \textbf{.0754} & \textbf{.1023} & \textbf{.0830} & \textbf{.1260} \\ \bottomrule
        \end{tabular}
        }
        \captionof{table}{Performance of SimPO and \ours's online DPO averaged across datasets.}
        \label{tab:po-algorithms}
    \end{minipage}\hfill
    \begin{minipage}{0.48\textwidth}
        \centering
        \includegraphics[trim=0 0.4cm 0 0.4cm, clip, width=0.95\linewidth]{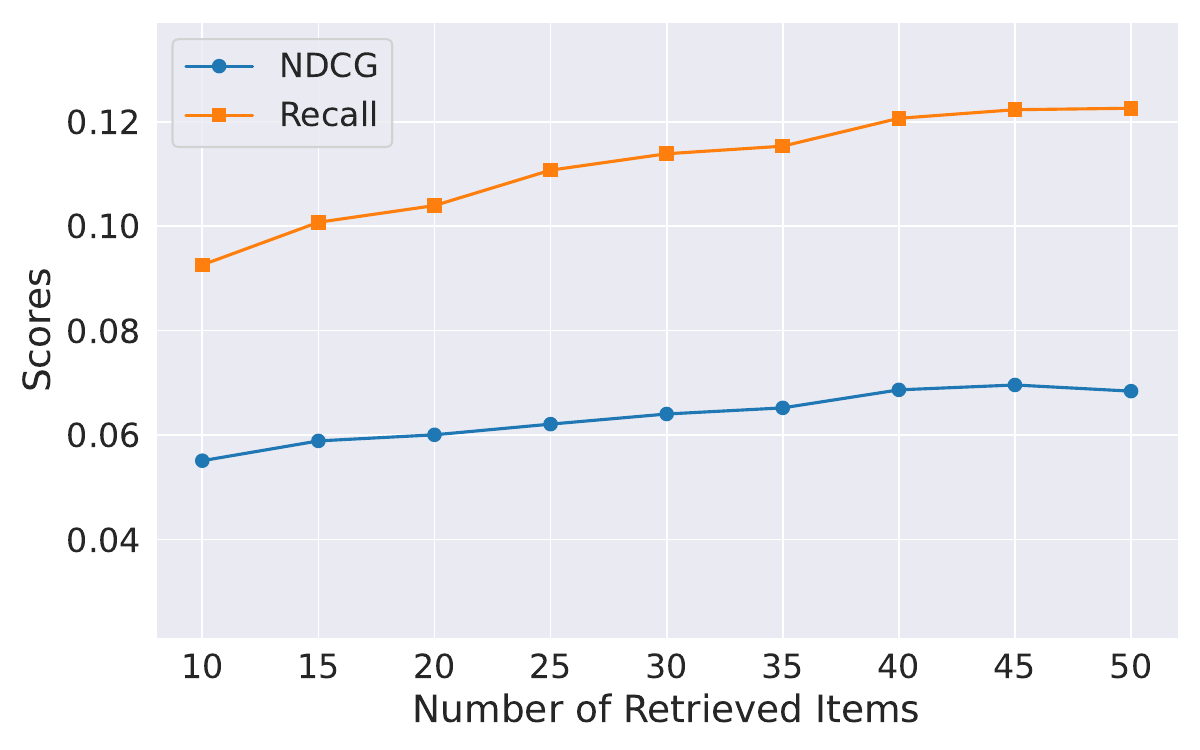}
        \captionof{figure}{Performance changes in \ours with different number of retrieved items.}
        \label{fig:num-items}
    \end{minipage}
    \vspace{-5pt}
\end{figure}

\noindent
\textbf{RQ2. Which retriever model works best?}
To analyze the retriever choice in \ours, we evaluate the performance of several different retriever models on the MovieLens dataset. In particular, we pretrain each retriever model and evaluate their retrieval performance on our collected corpus, with results presented in \Cref{tab:retriever}. Overall, LRURec achieves the highest scores across all reported metrics, recording an N@20 of 0.1700 and an R@20 of 0.3365, thereby outperforming all other retriever models. A similar trend is observed for the @10 metrics, where LRURec scores an N@10 of 0.1483 and an R@10 of 0.2508, establishing a consistent margin over other models. Among the baseline methods, FMLPRec serves as the strongest competitor, achieving the second-best results across most metrics (such as an R@20 of 0.3125 and an R@10 of 0.2411), followed closely by SASRec and BERT4Rec. Considering the efficiency advantages of LRURec in both training and inference, the findings indicate that LRURec is more effective at retrieving candidate items based on user history, leading to higher-quality retrieval outcomes for the following generation stage in \ours.

\noindent
\textbf{RQ3. How do different RL algorithms compare?}
We further analyze the core components of our online preference optimization by comparing two variants of our proposed method: the pairwise online DPO and the multi-sample GRPO (8 samples per example) using the Qwen backbone~\citep{shao2024deepseekmath}. The results are summarized in \Cref{tab:dpo-grpo}. As expected, we observe that our GRPO variant generally outperforms the pairwise DPO across most evaluation metrics and datasets. Specifically, the GRPO variant yields notable improvements on the N@5 metric, achieving 0.0385 compared to DPO's 0.0368 on Reddit. We attribute GRPO's superior performance to its multi-trajectory sampling, which yields robust relative advantage estimates for top-$k$ ranking. Due to GRPO's high inference costs and slower training, we adopt DPO as the default for \ours, which notably maintains 98.6\% of GRPO's average performance at a fraction of the computational cost. For DPO variants, we also adopt different formulations of \Cref{eq:dpo_loss} for comparison and present the average results across datasets in \Cref{tab:po-algorithms}. Here, we report SimPO (see details in \Cref{sec:app}) as an alternative~\citep{meng2025simpo}. Overall, \ours with DPO demonstrates consistently superior results compared to SimPO across all evaluated LLM backbones. For instance, \ours with Qwen achieves N@5 and R@5 scores of 0.0517 and 0.0740 respectively, outperforming SimPO on these crucial top-$k$ metrics. This dominating trend holds strictly true for both the Gemini and GPT backbones, where DPO consistently yields the higher performance across metrics. These findings validate that our proposed DPO-based preference optimization is highly effective at capturing nuanced user preferences, highlighting its potential for delivering higher-quality retrieval results in two-stage conversational recommendation.

\begin{figure}[t]
    \centering
    \begin{minipage}{0.49\textwidth}
        \centering
        \resizebox{0.9\linewidth}{!}{
        \small
        \begin{tabular}{@{}lcccc@{}}
        \toprule
        \multirow{2}{*}{$\beta$} & \multicolumn{4}{c}{\textbf{Average Performance}}                  \\ \cmidrule(l){2-5} 
                                 & N@5            & R@5            & N@10           & R@10           \\ \midrule
        0 (SFT)                  & .0469          & .0702          & .0515          & .0843          \\
        0.05                     & \textbf{.0513} & \textbf{.0741} & \textbf{.0580} & \textbf{.0953} \\
        0.1                      & .0482          & .0692          & .0547          & .0893          \\
        0.2                      & .0485          & .0711          & .0544          & .0888          \\ \bottomrule
        \end{tabular}
        }
        \captionof{table}{Recommendation performance of \ours with Qwen using different $\beta$ values.}
        \label{tab:beta}
    \end{minipage}\hfill
    \begin{minipage}{0.49\textwidth}
        \centering
        \includegraphics[trim=0 0 0 0, clip, width=0.9\linewidth]{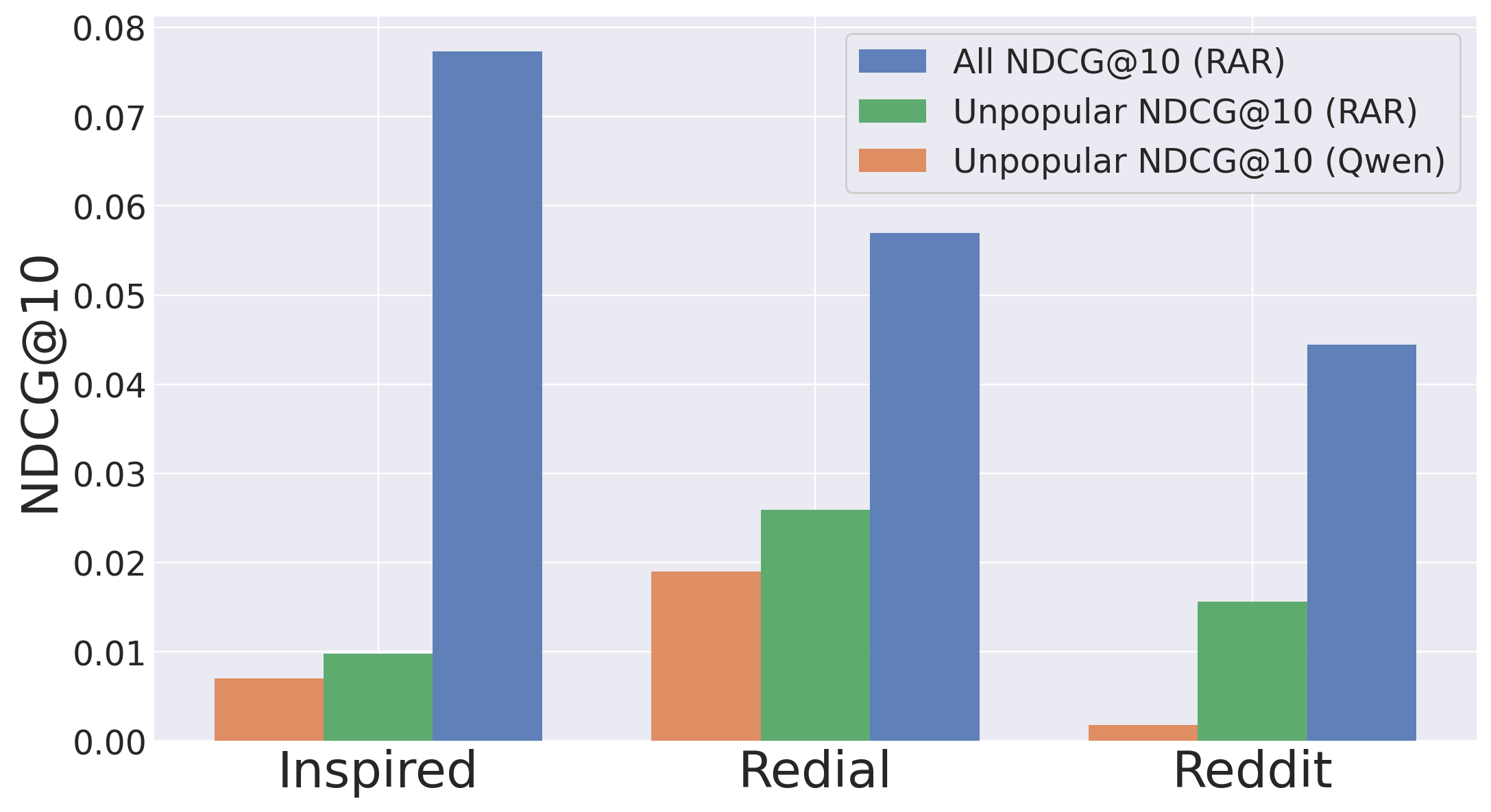}
        \captionof{figure}{N@10 on different item groups.}
        \label{fig:popularity}
    \end{minipage}
\end{figure}

\noindent
\textbf{RQ4. How do key hyperparameters impact the performance of \ours?}
Here, we investigate the impact of two critical hyperparameters on the performance of \ours: the number of retrieved items and the $\beta$ value.
First, we evaluate the performance changes when varying the number of retrieved candidate items, as illustrated in \Cref{fig:num-items}. The analysis reveals a clear, positive trend: as the number of retrieved items increases, both NDCG and Recall steadily improve. Recall naturally benefits from the expanded candidate pool, which increases the likelihood of capturing relevant items. More importantly, the ranking-sensitive metric, NDCG, also maintains an upward trajectory up to 45 items. This demonstrates the LLM generator's robust capacity to effectively filter and rank candidates without being immediately overwhelmed by noise. Furthermore, we examine the influence of the $\beta$ parameter, as summarized in \Cref{tab:beta}. Comparing various settings against the standard SFT baseline ($\beta=0$), we observe that introducing a small $\beta$ value significantly enhances recommendation accuracy. Specifically, setting $\beta=0.05$ achieves peak performance across all evaluated metrics, improving N@10 from $0.0515$ to $0.0580$ and R@10 from $0.0843$ to $0.0953$. Increasing $\beta$ further leads to a slight decline in performance; therefore, our experiments indicate that the optimal range for $\beta$ lies within $[0.05, 0.1]$. Overall, these observations suggest that \ours exhibits strong robustness across different hyperparameter choices, maintaining stable, high-quality recommendations without requiring exhaustive, fine-grained tuning.

\begin{table}[t]
\noindent\makebox[\textwidth]{
\resizebox{1.0\textwidth}{!}{
\begin{tabular}{@{}lcccccccccccc@{}}
\toprule
\multirow{2}{*}{Method} & \multicolumn{4}{c}{\textbf{Inspired}}                             & \multicolumn{4}{c}{\textbf{Redial}}                               & \multicolumn{4}{c}{\textbf{Reddit}}                               \\ \cmidrule(l){2-5} \cmidrule(l){6-9} \cmidrule(l){10-13}
                        & N@5            & R@5            & N@10           & R@10           & N@5            & R@5            & N@10           & R@10           & N@5            & R@5            & N@10           & R@10           \\ \midrule
Qwen$_{\text{on}}$      & \textbf{.0730} & \textbf{.0997} & .0764          & .1210          & .0438          & .0626          & .0501          & .0826          & .0367          & .0525          & .0402          & .0629          \\
Qwen$_{\text{off}}$     & .0693          & .0980          & \textbf{.0773} & \textbf{.1241} & \textbf{.0491} & \textbf{.0704} & \textbf{.0569} & \textbf{.0947} & \textbf{.0368} & \textbf{.0536} & \textbf{.0444} & \textbf{.0770} \\ \midrule
GPT$_{\text{on}}$       & .1006          & .1327          & .1115          & .1648          & \textbf{0.646} & \textbf{.0975} & \textbf{.0748} & \textbf{.1285} & \textbf{.0575} & \textbf{.0748} & \textbf{.0615} & \textbf{.0872} \\
GPT$_{\text{off}}$      & \textbf{.1091} & \textbf{.1422} & \textbf{.1180} & \textbf{.1700} & .0620          & .0932          & .0718          & .1236          & .0536          & .0711          & .0585          & .0859          \\ \bottomrule
\end{tabular}
}
}
\caption{Recommendation performance of LLMs with thinking enabled and disabled.}
\label{tab:reasoning-llms}
\vspace{-5pt}
\end{table}

\noindent
\textbf{RQ5. Does explicit reasoning improve LLM performance in CRS?}
To investigate whether extended reasoning inherently improve recommendation quality, we evaluated Qwen and GPT models with their explicit "thinking" capabilities enabled (on) and disabled (off). As detailed in \Cref{tab:reasoning-llms}, the impact of reasoning is not universally beneficial and varies significantly across scenarios. For the Qwen model, omitting the explicit reasoning phase consistently delivered competitive or superior performance across most datasets while significantly reducing inference costs. Conversely, GPT's behavior was highly dataset-dependent: $GPT_{\text{off}}$ achieved the highest metrics on the Inspired dataset, whereas $GPT_{\text{on}}$ performed strictly better on both Redial and Reddit. Ultimately, these findings indicate that forcing additional reasoning does not guarantee better recommendations; rather, its utility is dependent on both the underlying model and the specific data distribution.

\noindent
\textbf{RQ6. Does \ours improve popularity bias and hallucination?}
Finally, we analyze \ours (with the Qwen $f_{\mathrm{llm}}$) on two known LLM-based CRS challenges: popularity bias and item hallucination~\citep{he2023large, yang2024unleashing}. By categorizing items unseen in training into an unpopular group (\Cref{fig:popularity}), we confirm popularity bias exists; overall NDCG@10 scores substantially exceed those of unpopular items (e.g., Reddit scores just 0.0013). Nevertheless, \ours generally outperforms the LLM-only Qwen baseline. Leveraging a retriever with rich semantic embeddings and online RL-based post-training yields a nearly 4x improvement on unpopular items over Qwen. Additionally, we observe a dramatic reduction in hallucination rates, with under 1\% unmatched titles across all datasets. Ultimately, while not completely immune to popularity bias, \ours's retrieval augmentation design effectively mitigates hallucination, marking a critical step toward more reliable and trustworthy CRS.

%% file: latex/5_conclusion.tex
\section{Conclusion}
In this work, we introduce \ours, a two-stage framework for retrieval augmented conversational recommendation. For this purpose, we collect a unified text corpus of over 300k films within the movie domain to support our recommendation tasks. \ours employs a lightweight retriever to efficiently identify potential candidate items based on user interaction history. In addition, a black-box LLM is utilized to refine these candidates by integrating contextual data and retrieved information, thereby capturing subtle user preferences in natural language. To further align retrieval with generation, our online, on-policy preference optimization leverages LLM feedback to iteratively enhance the retriever's performance. Extensive experiments on benchmark datasets validate our approach, demonstrating that by effectively aligning the retriever and generator through online preference optimization, \ours consistently outperforms state-of-the-art baselines in conversational recommendation.

%% file: latex/6_appendix.tex
\newpage
\section{Appendix}
\label{sec:app}

\subsection{Corpus Collection}

\begin{table}[t]
\small
\centering
\resizebox{0.5\textwidth}{!}{
\begin{tabular}{@{}l|l@{}}
\toprule
\textbf{Attribute} & \textbf{Value}                      \\ \midrule
ID                 & \texttt{tt0111161}                  \\
Title              & The Shawshank Redemption            \\
Year               & 1994                                \\
Genre              & Drama                               \\
Director           & Frank Darabont                      \\
Cast               & Tim Robbins, Morgan Freeman ...    \\
Plot               & Chronicles the experiences of a ... \\ \bottomrule
\end{tabular}
}
\caption{An example movie with metadata in our corpus.}
\label{tab:metadata-example}
\end{table}

We report the details on the collection and processing of our unified movie text corpus. The process began by gathering an initial set of movie titles and identifiers (e.g., IMDb/TMDB) from multiple open-source datasets, including IMDb genre, IMDb media, and Inspired~\citep{hayati2020inspired, rajugc_imdbmovies, brightdata_imdbmedia}. We then augmented this initial set with movies mentioned in widely-used recommendation benchmarks, namely MovieLens, Redial and Reddit~\citep{harper2015movielens, li2018towards, he2023large}. A primary motivation for building a consolidated, offline corpus was to overcome the limitations of existing APIs. While services like TMDB\footnote{https://www.themoviedb.org/} provide extensive data, their APIs operate on a per-query basis with strict rate limits, making it infeasible to fetch metadata for all titles via individual HTTP requests. Furthermore, API responses are not directly linked to the specific movie mentions within our target conversational datasets. Our approach addresses these issues by creating a versioned, fully annotated dataset that can be indexed and processed in a single pass, thereby eliminating rate-limit bottlenecks and ensuring direct correspondence between metadata and conversational mentions. To construct this corpus, we first consolidated duplicated entries by verifying movie names and identifiers, resolving any conflicts by selecting the most informative metadata. For items with missing information, we carefully collected the missing values from TMDB or other online resources. We deliberately avoided direct scraping of copyrighted text fields from IMDb due to licensing restrictions\footnote{https://developer.imdb.com/}. Subsequently, we mapped the items in our corpus to their corresponding entries in the recommendation datasets (i.e., MovieLens, Inspired, Redial, and Reddit). This involved performing entity recognition and string matching to identify each movie mentioned in the conversational or sequential data and link it to its entry in our corpus. Any movies for which we could not find essential metadata (director, cast, genre, plot) were removed to maintain data quality. After processing, our final corpus contains 337,731 movie entries with comprehensive metadata, focusing primarily on English-language films (an example is shown in \Cref{tab:metadata-example}). The collection spans a wide temporal range, from films made as early as 1888 to upcoming releases in 2029, representing roughly half of all films listed on IMDb as of December 2024. By establishing explicit item correspondence to each of the benchmarks, the collected corpus serves as a robust and readily accessible knowledge base to train both the retriever and LLM components in \ours.

\subsection{Implementation \& Additional Results}
\label{sec:app2}

\begin{table}[t]
\small
\centering
\resizebox{0.5\textwidth}{!}{
\begin{tabular}{lrrr}
\toprule
Dataset   & \#Train & \#Val  & \#Test \\ \midrule
MovieLens & 536,127 & 67,015 & 67,015 \\
Inspired  & 1,507   & 206    & 183    \\
Redial    & 24,095  & 2,647  & 3,445  \\
Reddit    & 12,481  & 2,947  & 1,511  \\ \bottomrule
\end{tabular}
}
\caption{Dataset Statistics.}
\label{tab:statistics}
\end{table}

\textbf{Datasets.}
Our model is evaluated on three widely-used benchmark datasets for conversational recommendation: Inspired, Redial and Reddit~\citep{hayati2020inspired, li2018towards, he2023large}. To ensure the retriever has a robust initial understanding of the movie domain, we pretrain it on the MovieLens-20M dataset~\citep{harper2015movielens}. For the pretraining, we segment the dataset into short user sessions, using a 30-minute inactivity threshold to generate meaningful interaction sequences. During preprocessing, we construct input prompts with conversational context and incorporate additional retrieval augmentation from the collected corpus (see \Cref{sec:corpus}). For items retrieved during this process, we utilize a comprehensive set of attributes: \emph{title}, \emph{year}, \emph{genre}, \emph{director}, \emph{cast} and \emph{plot}. These attributes serve a critical dual purpose: they are first used to generate the initial item embeddings for training the retriever model, and subsequently, they function as structured context for the LLM generator. Detailed statistics for each dataset, including the number of training (\#Train), validation (\#Val) and test (\#Test) examples, are reported in \Cref{tab:statistics}.

\textbf{Baselines.}
For baseline models, we adopt \textit{text-based} SASRec, FMLP-Rec and LRURec following the fMRLRec implementation~\citep{kang2018self, zhou2022filter, yue2024linear, wang-etal-2024-train}. We also utilize \textit{supervised fine-tuning} (SFT) to learn the retriever model in combination with LLM as baselines. This approach aligns with the objective in sequential recommendation and maximizes the likelihood of ground truth items w.r.t. $\theta$. We report the details of baseline methods:
\begin{itemize}
    \item \emph{Knowledge-Based Recommender Dialog (KBRD)} integrates knowledge graph to understand conversational context and user preferences, enabling multi-turn reasoning over entities for recommendation~\citep{chen-etal-2019-towards}.
    \item \emph{Knowledge Graph-based Semantic Fusion (KGSF)} utilizes knowledge graph and a gated fusion mechanism to dynamically integrate semantic information from both conversation and item attributes~\citep{zhou2020improving}.
    \item \emph{Unified Conversational Recommender System (UniCRS)} presents a unified framework to handle diverse conversational goals, including recommendation and chitchat with a prompt-based approach via a shared encoder-decoder model~\citep{wang2022towards}.
    \item \emph{Self-Attentive Sequential Recommendation (SASRec)} is the first transformer-based sequential recommender. SASRec uses unidirectional self-attention to capture transition patterns~\citep{kang2018self}.
    \item \emph{Filter-enhanced MLP for Recommendation (FMLP-Rec)} also adopts an all-MLP architecture with filter-enhanced layers. FMLP-Rec applies fast Fourier transform to improve representation learning~\citep{zhou2022filter}.
    \item \emph{Linear Recurrence Units for Recommendation (LRURec)} is based on linear recurrence and is optimized for paralleled training. LRURec thus provides both efficient training and inference speed~\citep{yue2024linear}.
\end{itemize}
All models are implemented and trained according to the methodologies described in the original works, with unspecified hyperparameters used as recommended. For item encoding, we use the Qwen-8B embedding model and encode items with their available metadata in the format of key-value pairs~\citep{zhang2025qwen3}. For each recommender~/~retriever, we initialize with two layers and search the dropout rates among [0.2, 0.4]. The retriever models are pretrained on MovieLens 20M using an 8:1:1 train~/~validation~/~test split, where the data is split into sessions with a maximum time gap of 30 minutes. In pretraining, we sample 100 negative examples at each time step and utilize in-batch negatives to compute the negative log likelihood loss. The models with best validation performance are saved and evaluated on the test sets.

\begin{figure}[h]
\begin{prompt}{Example Prompt}

You are an expert in movie recommendations. Analyze the provided conversation history to identify the user's preferences, such as genres and actors. Then, rank the candidate movies by how well they match these preferences. Return your answer as a numbered list with each movie on a new line in the format: '<rank>. <movie name>'. Do not include any additional commentary, formatting or chattiness.
\vspace{3pt}

\texttt{<Retrieved Candidates w/ Metadata>}
\vspace{3pt}

Conversation history: 
\vspace{3pt}

\texttt{<Conversation Context>}
\vspace{3pt}

\end{prompt}
\caption{Example prompt for \ours. The prompt comprises of instructions, retrieved candidates, followed by the conversation context.}
\label{fig:prompt}
\end{figure}

\begin{algorithm*}[t]
\label{alg:rl}
\caption{Preference Optimization in \ours}
\begin{algorithmic}[1]
\Require
Conversational dataset $\mathcal{X}$, pretrained retriever $f_{\mathrm{ret}}$ (policy $\pi_{\theta}$), frozen LLM generator $f_{\mathrm{llm}}$, reward function $r(\cdot, \cdot)$, group size $g$ ($g=2$ for DPO), hyperparameters $\beta$, learning rate $\eta$.
\Ensure
Optimized retriever parameters $\theta$.

\State Initialize policy $\pi_{\theta}$ with parameters from pretrained $f_{\mathrm{ret}}$.
\State Initialize reference policy $\pi_{\mathrm{ref}} \leftarrow \pi_{\theta}$. \Comment{Copy initial weights}

\For{each training epoch}
    \For{each conversation $\mathcal{C} = (r_t, s_t, I_t)_{t=1}^{T}$ in $\mathcal{X}$}
        \For{$t = 1, \dots, T$}
            \If{$r_t$ is \emph{recommender}}
                \State Let history items be $\mathcal{I}_{\text{hist}} \leftarrow \{ I_{\tau} \}_{\tau=1}^{t-1}$.
                \State Let history sequence be $\mathcal{S}_{\text{hist}} \leftarrow \{ s_k \}_{k=1}^{t-1}$.
                
                \State Sample $\{C_i\}_{i=1}^g \sim \pi_{\theta}(\cdot | \mathcal{I}_{\text{hist}})$.
                
                \State Initialize rewards list $R \leftarrow []$.
                \For{$i = 1, \dots, g$}
                    \State Generate ranked list $\hat{I}_i \leftarrow f_{\mathrm{llm}}(\mathcal{S}_{\text{hist}}, C_i)$.
                    \State Calculate reward $r_i \leftarrow r(\hat{I}_i, I_t)$. \Comment{e.g., NDCG as reward}
                    \State Append $r_i$ to $R$.
                \EndFor
                
                \State Compute $\mathcal{L}_{\mathrm{rl}}$ with \Cref{eq:dpo_loss} or \Cref{eq:grpo_loss}.
                \State Compute NLL loss $\mathcal{L}_{\mathrm{nll}} \leftarrow -\log \pi_{\theta}(I_t | \mathcal{I}_{\text{hist}})$.
                \State Compute final loss $\mathcal{L} \leftarrow \mathcal{L}_{\mathrm{nll}} + \mathcal{L}_{\mathrm{rl}}$.
                \State Update parameters $\theta \leftarrow \theta - \eta \nabla_{\theta} \mathcal{L}$.
            \EndIf
        \EndFor
    \EndFor
\EndFor
\State \Return Optimized parameters $\theta$.
\end{algorithmic}
\end{algorithm*}

\begin{table*}[t]
\noindent\makebox[\textwidth]{
\resizebox{1.0\textwidth}{!}{
\begin{tabular}{@{}lcccccccccccc@{}}
\toprule
\multirow{2}{*}{Method} & \multicolumn{4}{c}{\textbf{Inspired}}                                 & \multicolumn{4}{c}{\textbf{Redial}}                                   & \multicolumn{4}{c}{\textbf{Reddit}}                      \\ \cmidrule(l){2-5} \cmidrule(l){6-9} \cmidrule(l){10-13} 
                        & N@5            & R@5            & N@10           & R@10           & N@5            & R@5            & N@10           & R@10           & N@5            & R@5            & N@10           & R@10           \\ \midrule
SFT$_{\text{Qwen}}$     & .0741          & .0869          & .0741          & .0869          & .0507          & .0736          & .0543          & .0843          & .0291          & .0410          & .0305          & .0450          \\
SFT$_{\text{GPT4o}}$    & .0547          & .0652          & .0600          & .0815          & .0514          & .0742          & .0585          & .0959          & .0340          & .0467          & .0371          & .0556          \\
SFT$_{\text{Gem2}}$     & .0793          & .0867          & \textbf{.1032} & .1250          &  \ul{.0588}    &  \ul{.0870}    &  \ul{.0684}    &  \ul{.1163}    &  \ul{.0418}    &  \ul{.0543}    &  \ul{.0443}    &  \ul{.0623}    \\ \midrule
RAR$_{\text{Qwen}}$     &  \ul{.0793}    &  \ul{.1032}    & .0793          & .1032          & .0540          & .0780          & .0583          & .0907          & .0326          & .0444          & .0330          & .0457          \\
RAR$_{\text{GPT4o}}$    & .0768          & .0978          & .0820          & .1141          & .0522          & .0771          & .0595          & .0994          & .0369          & .0510          & .0393          & .0582          \\
RAR$_{\text{Gem2}}$     & \textbf{.0831} & \textbf{.1087} &  \ul{.0995}    & \textbf{.1576} & \textbf{.0620} & \textbf{.0934} & \textbf{.0755} & \textbf{.1349} & \textbf{.0449} & \textbf{.0583} & \textbf{.0477} & \textbf{.0669} \\ \bottomrule
\end{tabular}
}
}
\caption{Additional results of \ours. For clarity, the best results for each dataset and metric are highlighted in \textbf{bold}, while the second-best results are \ul{underlined}.}
\label{tab:additional}
\end{table*}

For SFT baselines and \ours, we adopt LRURec as the retriever and further optimize the retriever model with the proposed online, on-policy preference optimization. The adopted LLMs are Qwen (Qwen3-8B), GPT-5 mini (gpt-5-mini-2025-08-07) and Gemini 3 Flash (gemini-3-flash-preview)~\citep{yang2025qwen3, singh2025openai}\footnote{https://deepmind.google/technologies/gemini/flash}. For training sets that are greater than 2.5k, we subsample to 2.5k for efficiency. Similarly, we search dropout rates among [0.1, 0.2]. Our learning rate is searched from [5e-5, 1e-4] and we adopt cosine scheduling for learning rate with warm-up steps of 100. SFT baseline models are trained without $\mathcal{L}_{\mathrm{dpo}}$ while in \ours, the $\beta$ values are searched within [0.05, 0.1, 0.2]. After sampling from the retriever, we annotate the candidate sets $C_1$ and $C_2$ using the LLM model. If any label item appears in both sets, we select the one with the higher label rank as $C_w$. If a label item is present in only one candidate set, that set becomes $C_w$. If neither set contains a label item, we resample until one of these conditions is met. Similar to baseline implementation, we keep models with best validation performance for evaluation. Our prompt is constructed as illustrated in \Cref{fig:prompt}. The LLM generation configuration is set to default, with the thinking efforts setting to \texttt{low} or \texttt{None} in our main experiments. All baseline methods and \ours are evaluated under identical conditions. Based on the LLM prediction, we perform string matching to compute the highest rank of label items, evaluation metrics are implemented following~\citep{he2023large}.

For alternative RL methods, we adopt the pairwise SimPO~\citep{meng2024simpo} and the multi-sample GRPO~\citep{shao2024deepseekmath}. GRPO details can be found in our main text, and the formulation for SimPO in our case can be formulated as~\citep{meng2025simpo}:
\begin{equation}
    \mathcal{L}_{\mathrm{simpo}} = -\log \sigma ( \beta \log P_{\theta} (C_w | \{ I_{\tau} \}_{\tau=1}^{t-1}) - \beta \log P_{\theta} (C_l | \{ I_{\tau} \}_{\tau=1}^{t-1}) - \gamma),
\end{equation}
where the the log probabilities are computed similar to \Cref{eq:likelihood}. Hyperparameters such as $\gamma$ are searched as recommended in the original paper. For GRPO, we set the group size to eight and compute the loss using \Cref{eq:grpo_loss}. Across all methods, we adhere to an online, on-policy setting to train and evaluate the retriever models. While our methodology formally introduces the reference model $\pi_{ref}$ in accordance with standard DPO, our final experiments omit this reference constraint. We empirically found that removing the reference model yields superior recommendation performance, a phenomenon consistent with recent findings in RLVR literature~\citep{yue2025hybrid, yu2025dapo}. Specifically, we omit the reference model for two primary reasons: (1) our joint optimization with the supervised negative log-likelihood loss ($\mathcal{L}_{nll}$) serves as a highly effective surrogate to prevent policy drift and maintain in-domain performance, rendering the KL penalty from a reference model largely redundant; and (2) while retriever models fundamentally require dropout to achieve optimal generalization, the inconsistency introduced by dropout destabilizes the log-likelihood calculations of the reference model, thereby failing to provide a stable or effective constraint on the policy. To compute hallucination rates, we perform fuzzy matching and consider a prediction to be hallucination if its similarity score to any of the provided candidate titles is below 0.85.

We present additional results to demonstrate that \ours generalizes well to other recent large language models. Specifically, we evaluate Qwen 2.5 (Qwen2.5-7B-Instruct), GPT-4o mini (gpt-4o-mini-2024-07-18), and Gemini 2.0 Flash (gemini-2.0-flash-001) \citep{yang2024qwen2, hurst2024gpt, comanici2025gemini}. The results are presented in \Cref{tab:additional}, from which we observe:
(1)~First, applying our method (RAR) consistently improves recommendation performance over the standard SFT baseline across all three foundation models. Whether using Qwen 2.5, GPT-4o mini or Gemini 2.0 Flash, the RAR variants yield higher NDCG and Recall scores in almost every scenario. (2)~the RAR formulation utilizing Gemini 2.0 Flash (RAR$_{\text{Gem2}}$) achieves the best overall results, securing the highest scores across all metrics for the Redial and Reddit datasets, as well as three out of four metrics for the Inspired dataset. 
In summary, these findings confirm that our approach is model-agnostic and generalizes highly effectively to various state-of-the-art large language models, reliably boosting performance beyond standard fine-tuning techniques.
